\documentclass[nonacm,sigplan]{acmart}

\newcommand{\ignore}[1]{}

\newcommand{\PG}{\mathcal{P}}
\newcommand{\CG}{\mathcal{C}}
\newcommand{\ham}{\mathcal{H}}
\newcommand{\CX}{CX}
\newcommand{\SWAP}{SW\!AP}
\newcommand{\loss}{\mathcal{L}}
\newcommand{\nairobi}{\texttt{nairobi}\,}
\newcommand{\toronto}{\texttt{toronto}\,}
\newcommand{\mumbai}{\texttt{mumbai}\,}
\newcommand{\hanoi}{\texttt{hanoi}\,}

\newcommand{\noise}{\mathcal{N}}
\newcommand{\GA}[1]{\ensuremath{\texttt{GA}_{#1}}}
\newcommand{\nstarts}{\texttt{s} }

\newcommand{\niter}{\texttt{m} }
\newcommand{\nbest}{\texttt{k} }

\usepackage[normalem]{ulem}
\usepackage{braket}
\usepackage{physics}
\usepackage{bbm}

\renewcommand{\vec}{\mathbf}

\usepackage{graphicx}
\graphicspath{figures/}

\begin{document}

\title{Clapton: Clifford Assisted Problem Transformation for Error Mitigation in Variational Quantum Algorithms}

\author{Lennart Maximilian Seifert} 
\email{lmseifert@uchicago.edu}
\affiliation{%
  \institution{University of Chicago}
  \city{Chicago}
  \state{IL}
  \country{USA}
}

\author{Siddharth Dangwal}
\email{siddharthdangwal@uchicago.edu}
\affiliation{%
  \institution{University of Chicago}
  \city{Chicago}
  \state{IL}
  \country{USA}
}

\author{Frederic T. Chong}
\email{chong@cs.uchicago.edu}
\affiliation{%
  \institution{University of Chicago}
  \city{Chicago}
  \state{IL}
  \country{USA}
}

\author{Gokul Subramanian Ravi}
\email{gsravi@umich.edu}
\affiliation{%
  \institution{University of Michigan}
  \city{Ann Arbor}
  \state{MI}
  \country{USA}
}

\begin{abstract}
Variational quantum algorithms (VQAs) show potential for quantum advantage in the near term of quantum computing, but demand a level of accuracy that surpasses the current capabilities of NISQ devices.
To systematically mitigate the impact of quantum device error on VQAs, we propose Clapton: \underline{Cl}ifford-\underline{A}ssisted \underline{P}roblem \underline{T}ransformati\underline{on} for Error Mitigation in Variational Quantum Algorithms. 
Clapton leverages classically estimated good quantum states for a given VQA problem, classical simulable models of device noise, and the variational principle for VQAs.
It applies transformations on the VQA problem's Hamiltonian to lower the energy estimates of known good VQA states in the presence of the modeled device noise.
The Clapton hypothesis is that as long as the known good states of the VQA problem are close to the problem's ideal ground state and the device noise modeling is reasonably accurate (both of which are generally true), then the Clapton transformation substantially decreases the impact of device noise on the ground state of the VQA problem, thereby increasing the accuracy of the VQA solution. 
Clapton is built as an end-to-end application-to-device framework and achieves mean VQA initialization improvements of 1.7x to 3.7x, and up to a maximum of 13.3x, over the state-of-the-art baseline when evaluated for a variety of scientific applications from physics and chemistry on noise models and real quantum devices.
\end{abstract}

\maketitle 
\pagestyle{plain} 

\section{Introduction}

Quantum computers are known to have the disruptive potential to provide a significant boost in computing capability, especially in important domains such as chemistry~\cite{kandala2017hardware}, optimization~\cite{moll2018quantum}, and machine learning~\cite{biamonte2017quantum}. They leverage unique principles of quantum mechanics like superposition, interference, and entanglement, which are beyond the capacity of classical computing, thereby enabling them to tackle problems that are considerably (sometimes exponentially) harder to solve on state-of-the-art classical computers.


In the near and intermediate future of quantum computing, we have to work with noisy intermediate-scale quantum (NISQ) devices~\cite{preskill2018quantum}. These devices are characterized by limited qubit quantity and quality. Their potential is constrained by issues like finite qubit coherence, imperfect state preparation and measurement operations (SPAM errors), gate errors and crosstalk. Therefore, they fall short when it comes to executing large-scale quantum algorithms that demand full fault-tolerance with error correction involving millions of qubits~\cite{O_Gorman_2017}. Despite these limitations, there is potential for near-term quantum advantage in certain tasks such as optimization~\cite{farhi2014quantum}, physics~\cite{kim_evidence_2023a}, and chemistry~\cite{peruzzo2014variational}. 
NISQ algorithms for these tasks have some inherent robustness to noise, which allows them to provide potentially useful results even without quantum error correction. 

An important family of such algorithms are variational quantum algorithms (VQAs). VQAs have a diverse range of applications such as estimating the electronic energy of molecules~\cite{peruzzo2014variational} and approximating MAXCUT problems~\cite{farhi2014quantum}. They are hybrid algorithms: The quantum circuits, which are executed on a quantum machine, are parameterized by a set of angles which is tuned by a classical optimizer. Over numerous iterations the VQA searches the quantum space to find the minimum ``energy'' of a target objective, which corresponds to the solution of the problem. The corresponding quantum state is called the problem's ``ground state''. 

However, even VQAs and other NISQ algorithms can sustain noise strengths only up to certain limits, and techniques that can reduce the effects of noise-induced errors are of paramount importance.
Multiple proposals of such \emph{error mitigation} methods have emerged over the past decade~\cite{czarnik2020error, Rosenberg2021, barron2020measurement, botelho2021error, wang2021error, takagi2021fundamental, temme2017error, li2017efficient, giurgica2020digital, ding2020systematic} and they have considerably improved the result quality of quantum algorithms.
However, there is still a long way to go -- despite the implementation of error mitigation techniques, the accuracy of VQAs on current NISQ machines often falls notably short of the stringent precision requirements essential in fields like molecular chemistry, particularly as the scale of problems grows.
Thus, novel error mitigation techniques tailored to VQAs can substantially enhance their real-world utility.

\begin{figure*}[htbp]
\centering
\includegraphics[width=0.9\textwidth]{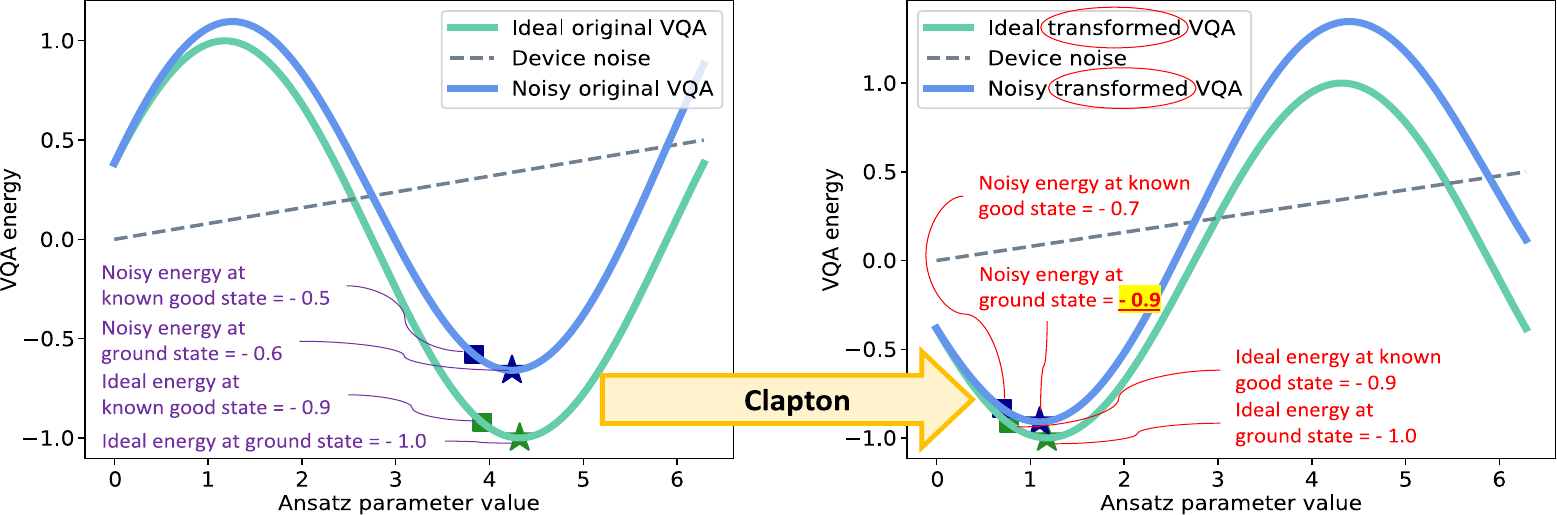}
\caption{The fundamental idea behind Clapton: Transforming a VQA problem structure to match the noise structure on a quantum device. Having identified a good algorithmic state, transforming the VQA problem such that it is mapped to a good hardware state improves the fidelity under noisy evaluation, leading to improved performance of the entire VQA instance.}
\label{fig:clapton_overview}
\end{figure*}

This work begins with the important observation that the noise affecting quantum devices has structure, stemming from physical properties of the qubits and their interaction with the environment.
The noise itself may vary temporally (over time) and spatially (across qubits), but a significant portion of the noise can be broken down into key components that can be modeled accurately and scalably with classical compute.
If such noise characteristics are well understood, then it is intuitive that quantum algorithms can be transformed so that they gain robustness to a target's device noise profile.
A simple example of a noise bias is the higher fidelity of the ground state $\ket{0}$ compared to the excited state $\ket{1}$ on most qubits due to state decay (more on this in Section \ref{subsec:nisq}). Thus, one could expect that quantum circuits that mostly operate on and navigate through quantum states with many excitations (large Hamming weight) are likely to show reduced fidelity compared to circuits which operate close to the $\ket{0}$ states throughout. Taking this a step further, if one has prior knowledge that a circuit is expected to produce outputs with large Hamming weight, and if they have the ability to efficiently transform this circuit to produce low Hamming weight outputs instead, then the execution fidelity of the transformed circuit is expected to be higher than the former. Applying the inverse transformation to the output should then yield a good solution to the original problem. 
This was the original motivation behind this work.

While state-dependent biases in quantum device noise and bias-aware circuit transformations are interesting, their utility is not obvious, because: (a) Useful quantum circuits produce output states unknown to the user, which is why the circuit is executed on the device in the first place. The classical computational complexity of identifying the correct output quantum state a priori would be exponential. (b) Naive classical simulation of the effects of device noise on a quantum circuit, to gauge its impact on a given quantum state, is exponentially hard as well. Thus, naively thinking, it would be implausible to use information about the correct output state and state-dependent noise biases to avoid the impact of detrimental execution on quantum devices.

However, there is unique opportunity in the domain of VQAs.
Specifically: (a) A VQA allows identifying noise-free ``good'' output states for the target problem (i.e., states having a high overlap with the correct output state of the circuit) which can be computed in polynomial time using classical methods. 
(b) A VQA can be efficiently transformed to an equivalent VQA problem with a different quantum state search space but an \emph{unchanged} the solution (the minimum energy), enabling a more noise-resilient search that is tailored to the quantum device characteristics. This is because there exists the freedom of a unitary basis change which only changes the representation of a problem (by defining new unit basis vectors in state space) but not its solution. As a fundamental operation in linear algebra this is universally applicable to any VQA problem.

Therefore, if we have a classically simulable model of some components of the device noise, 
then we are able to classically estimate the noisy fidelity of a known good state of a VQA problem and furthermore transform the VQA problem such that this state gets mapped to a good hardware state, like the all-qubit ground state $\ket{0}$, in a more noise-robust environment. This is the essence of our proposed work \emph{Clapton}: \emph{\underline{Cl}ifford-\underline{A}ssisted \underline{P}roblem \underline{T}ransformati\underline{on} for Error Mitigation in Variational Quantum Algorithms}. 
The Clapton hypothesis is that as long as the known good states of the VQA problem are fairly close to the ground state and the device noise modeling is reasonably accurate, our problem transformations are decreasing the impact of device noise on the ground state of the VQA problem, thereby increasing the accuracy of the VQA solution.
We design an end-to-end framework that implements this hypothesis and achieve tremendous improvements in VQA quality in noisy quantum environments.
Importantly, all the steps in the Clapton framework can be performed in polynomial time and are, therefore, classically efficient.
The high-level idea is illustrated in Figure \ref{fig:clapton_overview}.

We present the following key contributions and insights:
\begin{itemize}
    \item Introducing the software tool Clapton, which modifies the Hamiltonian of a VQA to match the noise characteristics of a target quantum device, enabling significantly improved accuracy of a VQA task.
    \item Demonstrating the importance of noise modeling when searching for good initializations to VQA problems on today's quantum devices.
    \item Evaluation of Clapton across a broad range of scientifically meaningful VQA problems, showing improvements for the VQA starting points and for parameter points after 300 VQA iterations, as well as improved VQA convergence properties.
    \item Demonstrating the impact of Clapton on real quantum hardware, showing that simplified, classically simulable noise models are sufficient to produce improvements in experiment.
    \item Presenting the influence of individual noise channels on VQA problems and how Clapton addresses them.
    \item Discussing the classical computational cost of Clapton and its dependance on problem size.
\end{itemize}

\begin{figure}[htbp]
\centering
\includegraphics[scale=0.9]{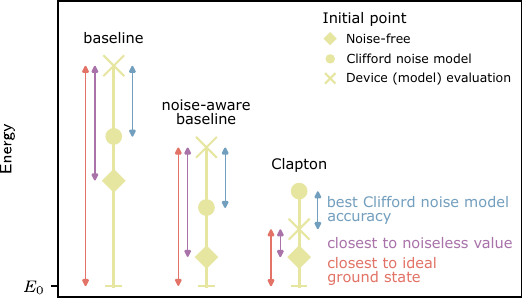}
\caption{\emph{Key result}: Clifford operations enable the combination of classically simulable noise and VQA Hamiltonian transformation, which ultimately gives Clapton its advantage over other initialization schemes in terms of solution quality (lower energy) and modeling accuracy.}
\label{fig:clapton_mini}
\end{figure}

Figure \ref{fig:clapton_mini} summarizes the key result of this work which will be motivated and developed throughout this manuscript.



\section{Background}\label{sec:background}

\subsection{Qubits, quantum gates and the Clifford group}\label{subsec:qubits_and_gates}
\label{subsec:qubits_gates_cliffs}
In quantum computing the fundamental unit of computation is the \emph{qubit} (quantum bit), which, contrary to classical bits, can exist in a superposition $\ket{\psi} = \alpha \ket{0} + \beta \ket{1}$ of its basis states. Similarly for $N$ qubits, their quantum state is given by a superposition of the $2^N$ basis states that correspond to the $2^N$ bitstrings $\qty{0,1}^N$. The quantum state is manipulated through the action of unitary matrices which are called \emph{quantum gates}. Typically the focus lies particularly on single- and two-qubit gates as they are sufficient to perform universal quantum computation \cite{nielsen2001quantum}. 

In the field of quantum information the set the single-qubit Pauli gates $\PG = \qty{I, X, Y, Z}$ is ubiquitous. The identity operation $I$ leaves the qubit state unchanged, $X$ causes a bit flip $\ket{0} \leftrightarrow \ket{1}$ and $Z$ flips the phase of the excited state, $Z\ket{1} = -\ket{1}$. As $Y = iXZ$, it can be considered as a combination of bit and phase flip up to a global phase $i$. The set of $N$-qubit Pauli operators is given by the $N$-fold tensor product $\PG_N = \PG^{\otimes N}$. Its elements are Pauli strings:
\begin{align}
    P = P^1 \otimes P^2 \otimes \dots \otimes P^N \equiv P^1 P^2 \dots P^N \in \PG_N.
\end{align}
$P^k$ denotes the Pauli operator acting on the $k$th qubit. Often $I$ components are not written explicitly, e.g. $IXIZ = X^2 Z^4$. 

Based on the definition of Pauli gates, another important group of quantum operators can be constructed. The \emph{Clifford group} $\CG_N$ on $N$ qubits includes all quantum gates $C$ that normalize the corresponding Pauli group $\PG_N$:
\begin{equation}
    \label{eq:clifford_group}
    C P C^\dagger \in \pm \PG_N \quad \forall \, P \in \PG_N.
\end{equation}
The expression $C P C^\dagger$ is called the \emph{conjugation} of the Pauli operator $P$ with the Clifford operator $C$ and by definition yields a Pauli operator again. However it can introduce a sign flip, indicated with $\pm$ here. We denote the conjugation of an input Pauli $P$ to an output Pauli $P'$ by $P \rightarrow \pm P'$. The Clifford group includes the set of Pauli gates, $\PG_N \subset \CG_N$. 

In this work we primarily focus on single- and two-qubit Clifford gates. We emphasize the conjugation effect of the two-qubit controlled-NOT gate $\CX_{c \rightarrow t}$ with control on qubit $c$ and target on qubit $t$:
\begin{equation}
    \label{eq:cx}
    \begin{aligned}
        X^c &\rightarrow X^c X^t, &\qquad X^t &\rightarrow X^t, \\
        Z^c &\rightarrow Z^c, &\qquad Z^t &\rightarrow Z^c Z^t.
    \end{aligned}
\end{equation}
It is important to note that it correlates certain Pauli terms with each other and can therefore influence the way errors propagate through quantum circuits.


\subsection{Noisy Intermediate-Scale Quantum Computing}
\label{subsec:nisq}
Today's quantum devices are noisy~\cite{preskill2018quantum}, caused by unwanted interactions with the environment and imperfect control of the quantum system. Currently, noise in hardware is strong enough that it renders most computations meaningless. For example, the average error rate of a two-qubit gate is about 1\% on existing hardware from IBM and Google, whereas that of measurement operations is about 4\%~\cite{sycamoredatasheet}, limiting the fidelity of quantum programs on NISQ devices. We describe some prominent sources of error in the following.

\subsubsection{Thermal relaxation}
A quantum system has a natural tendency to decay to a low energy configuration. Thus, a qubit initialized in the arbitrary superposition state $\alpha\ket{0} + \beta\ket{1}$ progressively decays down to the $\ket{0}$ state as it evolves over time. The probability of decaying from the $\ket{1}$ to the $\ket{0}$ state after a time $t$ is given by $e^{\frac{-t}{T_1}}$, where $T_1$ is the characteristic decay time of a qubit describing its thermal relaxation.

\subsubsection{Gate errors}
Quantum states are acted upon by quantum gates to achieve a desired target state, i.e. $\ket{\psi} \rightarrow U\ket{\psi}$. However, quantum gates $U$ on NISQ devices are noisy. A good way to model this effect is given by the noiseless application of $U$ followed by a \textit{depolarizing noise channel}, which is realized by the stochastic application of a random Pauli gate. In particular, gate error rates gained from experimental methods like randomized benchmarking correspond to depolarizing error probabilities \cite{magesan2011rb}.
    
\subsubsection{Measurement errors}
When a qubit is measured, it collapses from a superposition state to one of the basis states, i.e. $\alpha\ket{0} + \beta\ket{1} \rightarrow \ket{0}/\ket{1}$. However, the qubit might suffer errors at the time of measurement due to imperfect measurement control or misclassification of the result, i.e. $\ket{0}$ is reported as $\ket{1}$, and vice versa. 

\subsection{Variational quantum algorithms}\label{subsec:vqa}
Despite these errors, there are certain classes of algorithms where current noisy quantum computers are postulated to outperform the most powerful classical computers. One such class is variational quantum algorithms (VQAs).

\begin{figure}[htbp]
    \centering
    \includegraphics[width=0.8\linewidth]{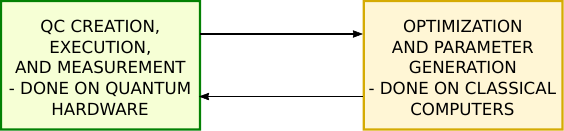}
    \caption{VQAs combine a classical optimizer which adjusts the parameters in each iteration with a quantum device producing function evaluations by running the corresponding quantum circuits.}
    \label{fig:vqe_background}
\end{figure}

A VQA consists of a parameterized circuit, called an \emph{ansatz}, and a Hermitian operator of interest $\ham$, called its \emph{Hamiltonian}. The ansatz is usually constructed from parameterized single-qubit rotations interspersed with entangling gates and it prepares a quantum state depending on those parameters. A scalar loss (objective function) is given by the expectation value $\ev{\ham}$ (``energy'') of the problem Hamiltonian for this quantum state, which can be obtained by measurement on quantum hardware. The goal is to obtain an optimal set of parameters that minimizes the loss. This is achieved by employing a classical minimization algorithm which feeds ansatz parameters to the quantum device and processes the resulting loss. Gradient-based methods like SPSA and gradient-free methods like Nelder-Mead have been used for this task in the past \cite{TILLY20221}. Figure \ref{fig:vqe_background} summarizes the process.

VQAs exhibit \textit{Optimal Parameter Resilience (OPR)} \cite{wang2021can}: If the noise is below a certain threshold, the parameters that minimize the loss on noisy hardware also often minimize the loss on a noiseless device. This makes VQAs robust to errors and leading candidates for demonstrating quantum advantage on NISQ devices. However, often the noise on NISQ devices is too high to observe OPR. Thus, we need to suppress errors using error mitigation techniques tailored for VQAs in order to realize OPR.

\subsection{Variational Quantum Eigensolver (VQE)}\label{subsec:vqe}
A popular VQA, which involves the estimation of the minimum eigenvalue of a given Hamiltonian $\ham$, is called the Variational Quantum Eigensolver (VQE) \cite{TILLY20221}. VQE is extensively used in physics and chemistry applications for finding the minimum energy of Ising Models or computing dissociation energies of molecules as functions of their bond lengths. This is why VQE has been studied extensively from an architectural perspective \cite{ravi2022cafqa, adaptvqe, wang2022quantumnas, ravi2021vaqem} to ensure its reliable execution. In this paper we specifically focus on VQE to benchmark Clapton, however it is important to note that Clapton can be applied to other VQAs as well.

\subsection{CAFQA}
For VQAs, the optimal circuit parameters and the target objective is, of course, unknown. However, prior work has proposed effective classical methods to find good initial circuit parameters for VQA problems~\cite{hartree1935self,ravi2022cafqa}.
Of particular interest to this work is CAFQA~\cite{ravi2022cafqa}, which uses classical noiseless simulation to explore the Clifford space of a target VQA problem to find a good initial state.
CAFQA exploits the key idea that circuits made up of only Clifford operations can be exactly simulated in polynomial time~\cite{gottesman1998heisenberg}, which is due to their special properties mentioned in Section \ref{subsec:qubits_gates_cliffs}.

Clifford operations do not provide a universal set of quantum gates - hence, the so-called stabilizer states produced by Clifford-only circuits are limited in how effectively they can explore the quantum space. However, CAFQA showed that the Clifford-based initial state has high overlap with the problem's ground state, and often achieved 90-99\% accuracy in the finding the ground state energy, even before running VQA on the actual quantum device. While CAFQA only targeted VQA initialization, our work Clapton pursues the idea that these known good initial states provide excellent opportunity for the state-dependent error mitigation.

CAFQA's main shortcoming is that it does not take noise characteristics of the underlying hardware into account, which makes it susceptible to device errors that can significantly degrade the quality of the found solution when executed on a quantum machine, therefore not guaranteeing high-fidelity results in experiment.


\ignore{
Variational algorithms expect to have innate error resilience due to hybrid alternation with a noise-robust classical optimizer~\cite{peruzzo2014variational, mcclean2016theory}. 
An overview of this process is illustrated in Fig.~\ref{fig:vqa}.
There are multiple applications in the VQA domain such as the Quantum Approximate Optimization Algorithm (QAOA)~\cite{farhi2014quantum} and the Variational Quantum Eigensolver (VQE)~\cite{peruzzo2014variational}.
{Our applications in this work target VQE, but VarSaw is applicable to all VQA problems - more on this in Section \ref{FW}.
VQE in itself has wide applications in  molecular chemistry, condensed matter physics, quantum Ising optimization problems, and a variety of quantum mechanics many-body problems, etc.}

An important application of VQE is the ground state energy estimation of a molecule, a task that is exponentially difficult in general for a classical computer~\cite{Gokhale:2019}.
Estimating the molecular ground state has important applications in chemistry, such as determining reaction rates and molecular geometry.

\rev{The quantum circuit used in each iteration of VQE (and VQA in general) is termed an ansatz which describes the range of valid physical systems that can be explored and thus determines the optimization surface. Traditionally, the ansatz is parameterized by 1-qubit rotation gates.
A good ansatz provides a balance between a simple representation, efficient use of available native hardware gates, and sufficient sensitivity with the input parameters.
The classical tuner/optimizer~\cite{9259985,SPSA} variationally updates (often via a gradient based approach) the parameterized circuit until the measured objective converges to a minimum.} 
An example 3-qubit ansatz (i.e., the paramerized circuit which is tuned in VQA) is shown in Fig.\ref{fig:jig_pauli_ansatz} - { measurements have to be} performed on this ansatz every iteration, which is key to the motivation for this work (more on this later).

\begin{figure}[t]
\centering
\includegraphics[width=0.98\columnwidth,trim={0.1cm 0cm 0cm 0cm},clip]{figures/varsaw_ansatz.pdf}
\caption{
Ansatz circuit measured on different Pauli bases (left: `ZZZ', right: `XZX'). Qubit commutativity allows for a set of Pauli strings to be measured on a single basis.
}
\label{fig:jig_pauli_ansatz}
\end{figure}

Estimating the VQE global optimum with high accuracy has proven challenging in the NISQ era even with sophisticated optimizers, a well-chosen ansatz, and error mitigation~\cite{ravi2021vaqem,czarnik2020error,Rosenberg2021,barron2020measurement,botelho2021error,wang2021error,tilly2021variational,takagi2021fundamental}. 
\iffalse 
As an example, ground state energy estimation of molecules (the energy required to break a molecule into its sub-atomic components), a key use case for VQE, requires energy estimates with an estimation error of less than $1.6 \times 10^{-3}$ Hartree, or what is known as ``chemical accuracy"~\cite{peterson2012chemical}, for applicability in understanding chemical reactions and their rates.
Unfortunately, for instance, prior work on the estimation of ground state energy of BeH$_2$ on a superconducting transmon machine resulted in an error greater than 10$^{-1}$ Hartree, which is roughly 100x worse than the required accuracy~\cite{kandala2017hardware}. 
}

\ignore{

\section{Quantum computing formalism}

In the field of quantum information a special set of gates are the single-qubit Pauli operators $\PG^{(1)} = \qty{I, X, Y, Z}$. Here we explicitly include the identity $I$ so that $\PG$ forms a basis of all single-qubit operators. 

Due its exponentially-scaling nature, quantum computation is generally infeasible to simulate. However there exists a subset of states, called \emph{stabilizer states}, that can be simulated in polynomial times due to an efficient representation in terms of Pauli strings. A $N$-qubit stabilizer state $\ket{\psi}$ is the mutual $+1$ eigenstate of $N$ commuting  \emph{stabilizers} $\mathcal{S} = \qty{S_i}_{i=1}^N$:
\begin{align}
    S_i \ket{\psi} = \ket{\psi} \quad \forall \, i=1, \dots, N
\end{align}
Stabilizers are Pauli strings $S_i = \qty(P^1 P^2 \dots P^N)_i$ and specifying the stabilizer set $\mathcal{S}$ uniquely fixes the corresponding state $\ket{\psi}$ (up to global phase). This representation grows linearly in the number of qubits which essentially is the foundation of the exponential speedup simulations benefit from when limiting to this state space.

}


\section{Clapton: Motivation and proposal}
Clapton addresses these issues by incorporating hardware information from the lowest layer of the stack and modeling relevant sources of noise in Clifford-based efficient simulations. Given a VQE problem $\ham$, it searches for a transformation $\hat{\ham} = \hat{C}^\dagger \ham \hat{C}$ to an equivalent problem $\hat\ham$ with improved resilience to realistic hardware errors. $\ham$ and $\hat\ham$ are equivalent formulations of the same optimization problem because they share the same eigenvalues (in particular, the ground state energy which is given by the smallest eigenvalue) due to the unitary transformation via the Clifford operation $\hat{C}$.

\subsection{The big picture of the Clapton transformation}
Let us consider a VQE problem $\ham$ on $N$ qubits with a variational ansatz $A(\theta)$ over $d$ rotation parameters $\theta \in [0, 2\pi)^d$. The VQE is solved by finding the minimum expectation value over all quantum states that can be produced by the ansatz:
\begin{equation}
      E^* = \min_\theta \ev{\ham}{\psi(\theta)} = \min_{\theta} \ev{A^\dagger(\theta) \ham A(\theta)}{0}.
\end{equation}
Here $E^* \geq E_0$ denotes the best approximation to the true ground state energy $E_0$ achievable in the variational setting. CAFQA limits the search over discrete sets of parameters $\theta$ such that $A(\theta)$ becomes a Clifford operation and is efficiently simulable on classical hardware.

Clapton changes perspective by searching over Hamiltonian transformations $\ham(\gamma) = C^\dagger(\gamma) \ham C(\gamma)$ where the qubit ground state $\ket{0}^{\otimes N} \equiv \ket{0}$ corresponds to the best initial state for the VQE problem. The full "post-Clapton" VQE algorithm on the quantum device is then initialized with ansatz parameters $\theta = \vec{0}$. Clapton finds this transformation by primarily focusing on the following optimization problem:
\begin{equation}
    \begin{aligned}
        \hat{\gamma} &= \arg \min_\gamma \ev{\tilde{A}^\dagger(\vec{0}) \ham(\gamma) \tilde{A}(\vec{0})}{0} \\
        &= \arg \min_\gamma \ev{\tilde{A}^\dagger(\vec{0}) C^\dagger(\gamma) \ham C(\gamma) \tilde{A}(\vec{0})}{0}. 
    \end{aligned}
\end{equation}
$C(\gamma)$ corresponds to a parametrized Clifford operation with parameters $\gamma \in \Gamma$. The set of Clifford operations on $N$ qubits scales exponentially with $N$ \cite{selinger_generators_2015}, therefore it is useful to choose an ansatz for $C$ as well to limit the search space. $\tilde{A}$ denotes the noisy version of the ansatz $A$. We model this noise with stochastic Pauli operations (see Section \ref{subsec:clapton_noise}). The goal is to resemble the hardware noise as accurately as possible to account for its effect during the online VQE optimization that follows Clapton. However, it is crucial to note that this design choice makes $\tilde{A}(0)$ a composition of (in part stochastic) Clifford operations, meaning it can be efficiently evaluated with classical compute. Clapton's output is the transformation parameter vector $\hat\gamma$ which yields the transformation operator $\hat{C} = C(\hat\gamma)$ and new problem formulation $\hat\ham = \ham(\hat\gamma)$.

\subsection{Transformation via Pauli conjugation}
\label{subsec:transformation}
A VQE Hamiltonian is typically expressed as a weighted sum of $M$ Pauli terms $\ham = \sum_{i=1}^M c_i P_i$, where $c_i$ is the energy coefficient corresponding to the Pauli string $P_i$. Clapton's transformation with a Clifford operation $C$ maps every Pauli string $P_i$ to its \emph{anti}conjugated Pauli string $C^\dagger P_i C = \pm P_i'$. The sign is absorbed into the coefficient $c_i \rightarrow c_i' = \pm c_i$, therefore the transformed Hamiltonian reads:
\begin{equation}
    C^\dagger \ham C = \sum_{i=1}^M c_i' P_i'.
\end{equation}
This shows one of the appealing properties of Clifford gates: Since they map Pauli strings to (signed) Pauli strings, the structure of the VQE problem is preserved and the transformed Hamiltonian can be implemented in the VQE framework just like the original one.

Anticonjugation is the inverse operation compared to the formal definition introduced in Equation \eqref{eq:clifford_group}. The choice of conjugation method is arbitrary because $C^\dagger$ is Clifford if and only if $C$ is, so the same properties hold. We prefer the anticonjugation because it is more intuitive in the grand scheme of VQE: When running the post-Clapton VQE algorithm, a quantum state $\ket*{\hat\psi(\theta)} = A(\theta) \ket{0}$ with respect to the Clapton Hamiltonian $\hat\ham$ can be directly translated back to a corresponding quantum state $\ket{\psi(\theta)} = \hat{C} \ket*{\hat\psi(\theta)} = \hat{C} A(\theta) \ket{0}$ with respect to the original Hamiltonian $\ham$. The states are equivalent as far as VQE is concerned because they share the same energy, $\ev*{\ham}{\psi(\theta)} = \ev*{\hat{C}^\dagger \ham \hat{C}}{\psi(\theta)} = \ev*{\hat\ham}{\hat\psi(\theta)}$. Therefore $\hat{C}$ can be used to construct a good solution with respect to the original Hamiltonian, which is particularly easy in experiment if $\hat{C}$ is given by a circuit using single- and two-qubit Clifford gates like we are suggesting.

\section{Clapton design}
\begin{figure*}[htbp]
    \centering
    \includegraphics[width=\linewidth]{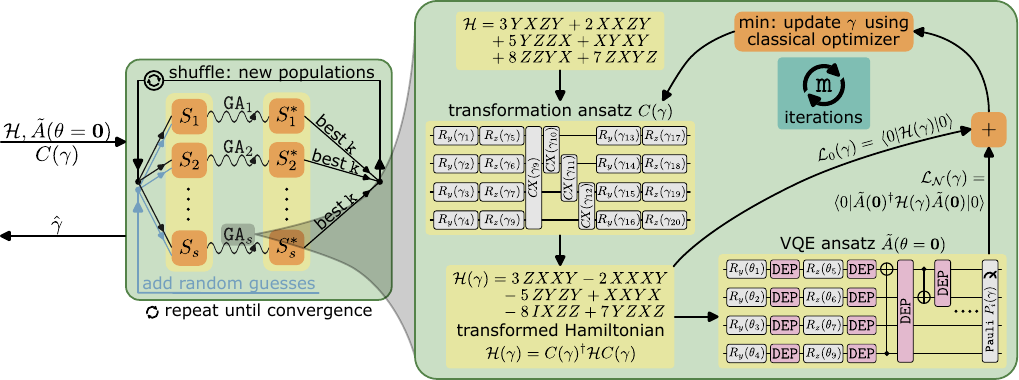}
    \caption{The Clapton transformation engine. Given a problem Hamiltonian $\ham$, a noise-equipped version of the VQE ansatz (at the zero point) $\tilde{A}(\vec{0})$ and a Clifford transformation ansatz $C(\gamma)$, Clapton employs genetic algorithms (GAs) to search over transformation parameters $\gamma$ to find a solution $\hat{\gamma}$. It instantiates \nstarts processes $\GA{1}, \dots, \GA{s}$, starting from random popultions $S_1, \dots, S_s$, that search the parameter space independently. Internally, each \GA{} instance performs \niter iterations of transformation optimization, where for different $\gamma$ the Hamiltonian is transformed and the noisy ($\loss_\noise$) as well as noiseless ($\loss_0$) energy of the initial point is evaluated. The GAs produce final populations $S_i^*$, of which Clapton takes the top \nbest solutions, mixes them and randomly constructs new starting populations $S_i$, additionally adding new random guesses that are not derived from the previous round. This entire scheme is repeated until the global loss value does not decrease between rounds (convergence).}
    \label{fig:clapton}
\end{figure*}

Within the scope of this paper we focus on specific choices for the VQE ansatz $A(\theta)$ and the Clapton transformation ansatz $C(\gamma)$. For the VQE ansatz we consider the circular hardware-efficient ansatz. It is composed of parameterized rotation gates $R_y$ and $R_z$, where
\begin{equation}
    R_y(\theta_j) = \exp(-i \frac{\theta_j}{2} Y), \quad R_z(\theta_j) = \exp(-i \frac{\theta_j}{2} Z),
\end{equation}
as well as controlled-NOT gates ($\CX$). The number of parameters is $d=4N$. If $\theta \in \qty{0, \pi/2, \pi, 3\pi/2}^d$, then all rotation gates are Clifford, turning the full ansatz circuit into an efficiently simulable Clifford operation. CAFQA searches over this discrete space directly to find the best Clifford initialization for $A(\theta)$ in noiseless simulation. For Clapton we are especially interested in the circuit $A(\vec{0})$, where all the rotation gates become $I$ gates and only the $\CX$ skeleton remains. 

The transformation ansatz $C(\gamma)$ is inspired by the VQE ansatz, however it treats the two-qubit gates differently. While it still includes the parameterized rotation gates, it replaces the $\CX$ operations with parameterized two-qubit gates, which, depending on the added parameters $\gamma_j$, can each be one of the following (acting on qubits $k$ and $l$):
\begin{equation}
    \begin{aligned}
        \gamma_j = 0&: I^k I^l, &\qquad \gamma_j = 1 &: \CX_{k \rightarrow l}, \\
        \gamma_j = 2&: \CX_{l \rightarrow k}, &\qquad \gamma_j = 3 &: \SWAP_{k \leftrightarrow l}.
    \end{aligned}
\end{equation}
We do this to increase the expressiveness of the ansatz and include the possibility to move around the components $P_i^k$ of the Hamiltonian Pauli strings $P_i$ by the means of $\SWAP$ operations ($\gamma_j = 3$). Furthermore, given the properties of $\CX$ that were discussed after Equation \eqref{eq:cx}, having a $CX$ ($\gamma_j = 1,2$) or not ($\gamma_j = 0$) as well as the choice of direction ($k \rightarrow l$ versus $l \rightarrow k$) has a significant impact on the transformed Hamiltonian and its characteristics in a noisy environment. Our Clapton transformation ansatz for four qubits is visualized as part of the right flowchart shown in Figure \ref{fig:clapton}.

Clapton's parameter space $\Gamma$ for this ansatz has dimension $5N$ and each parameter $\gamma_j$ can take four values, either one of four rotation angles for a single-qubit gate or one of the four choices for a two-qubit gate.

\subsection{The implementation of the Clapton optimization}
Just like CAFQA, Clapton focuses on finding a good initial point for a given VQE problem. Unlike CAFQA, it achieves this by transforming the Hamiltonian of the problem and restricting the initial parameters for the post-Clapton VQE to be $\theta = \vec{0}$. Thus, one part of Clapton's cost function is given by the noisy estimate of the energy of the transformed Hamiltonian $\ham(\gamma)$ at the initial VQE point:
\begin{equation}
    \label{eq:loss_n}
    \loss_\noise(\gamma) = \ev*{\tilde{A}^\dagger(\vec{0}) \ham(\gamma) \tilde{A}(\vec{0})}{0}. 
\end{equation}
However, only searching for a low noisy energy is not sufficient. There can be solutions that, when evaluated without noise, are relatively far from the true ground state energy $E_0$ but they are also rather error-resilient, therefore their noisy evaluations appear deceptively good. Thus a transformation $\gamma$ will be useful if it shows low noisy \emph{and} noiseless energy. This is why Clapton furthermore adds the noiseless energy as another term
\begin{equation}
    \loss_0(\gamma) = \ev*{A^\dagger(\vec{0}) \ham(\gamma) A(\vec{0})}{0} = \ev*{\ham(\gamma)}{0}
\end{equation}
($A(\vec{0})\ket{0} = \ket{0}$, see Section \ref{subsubsec:thermal}), so that the full cost function reads $\loss(\gamma) = \loss_\noise(\gamma) + \loss_0(\gamma)$.

Clapton evaluates both terms using the open-source framework stim \cite{gidney2021stim} that has been developed for fast Clifford simulations. It provides functionality to compute the anticonjugation of Pauli strings under Clifford operations as well as tools to simulate noisy Clifford circuits with depolarizing errors. Clapton then solves the discrete optimization problem 
\begin{equation}
    \hat{\gamma} = \arg \min_\gamma \loss(\gamma)
\end{equation}
by searching the parameter space $\Gamma$ using genetic algorithms (GAs) implemented with PyGAD \cite{gad2021pygad}.

The complete optimization engine is visualized in Figure \ref{fig:clapton}. Feeding in the VQE problem ($\ham$) and setup ($\tilde{A}(\vec{0})$, $C(\gamma)$), Clapton spawns \nstarts GA instances that individually search $\Gamma$ starting from random populations $S_i$. They each compute \niter iterations using the loss function $\loss(\gamma)$, yielding final populations $S_i^*$. Now Clapton combines good features from different GA instances. Therefore it chooses the best \nbest solutions from each $S_i^*$, mixes them and randomly draws \nstarts new initial populations, while adding new random guesses to fill back up to the original population size $\abs{S}$. This is repeated until the global loss does not decrease between rounds, where Clapton typically allows for two retry rounds in this case before terminating. For this work we ran  Clapton with the hyperparameters $\nstarts = 10$, $\niter = 100$, $\nbest = 20$ and $\abs{S} = 100$, which we found to work consistently well.

\subsection{Clapton is designed to account for realistic noise}
\label{subsec:clapton_noise}
Here we describe how the Clapton design addresses the three relevant sources of error  introduced in Section \ref{subsec:nisq}.

\subsubsection{Thermal relaxation}
\label{subsubsec:thermal}
As this error channel causes excitations ($\ket{1}$) to decay, the qubit ground state $\ket{0}$ is the favored state in its presence. This is why Clapton chooses the initial point $\theta = \vec{0}$ for the post-Clapton VQE: In the idealized case of no gate errors, the VQE circuit $A(\vec{0})$ keeps all qubits in their ground state because the rotation gates reduce to identity operations and the $\CX$ gates do not have an effect, $A(\vec{0}) \ket{0} = \ket{0}$. Furthermore, recalling the discussion from the end of Section \ref{subsec:transformation}, the ground state $\ket{0}$ translates to the equivalent state $\ket*{\hat{\psi}} = \hat{C} \ket{0}$ with respect to the original Hamiltonian $\ham$. The work on CAFQA has shown that Clifford states like $\ket*{\hat{\psi}}$ can 
approximate the true solution of $\ham$ very well. This implies that $\ket{0}$ represents a good approximation to the solution of $\hat\ham$ as well. Thus it is reasonable to expect that the true solution of $\hat\ham$ shows a large overlap with the ground state and the measurement results will be dominated by the ground state during the VQE execution. Therefore, by prioritizing the $\ket{0}$ state this design implicitly leads to increased resilience to thermal relaxation when performing VQE on the transformed problem $\hat\ham$.

\subsubsection{Gate errors}
However, in reality quantum instructions are imperfect and gate errors must be considered. This requires replacing the ansatz circuit $A(\vec{0})$ by its noisy version $\tilde{A}(\vec{0})$. As mentioned previously, a common way to model gate errors is with depolarizing noise. A single-qubit gate error of strength $p$ is then realized by the application of $X$, $Y$ or $Z$ with chance $p/3$ after the gate. Similarly, a two-qubit gate error of strength $p$ is implemented by appending one of the 15 two-qubit Pauli strings (excluding $I^{\otimes 2})$ with chance $p/15$. This is the convention used in stim but we note that other conventions for these parameters exist. Each time the circuit is simulated (a shot is executed) the random gates get sampled which ultimately leads to the noisy output distribution of the circuit. As $\tilde{A}(\vec{0})$ is only constructed from (in part stochastic) Clifford operations, it is efficiently simulable in stim. By incorporating gate errors Clapton becomes aware of which qubits might accumulate a significant amount of noise and adjust the Hamiltonian accordingly to minimize the effect on the evaluation of Pauli terms with large energy coefficients. In particular, $\CX$ gates propagate and spread errors due to their special nature (see Equation \eqref{eq:cx}), which especially affects qubits at the end of long $\CX$ chains.

\subsubsection{Measurement errors}
\label{subsubsec:meas_errors}
Measurement is typically performed in the computational basis and yields classical bit values $0$ or $1$ for each measured qubit as output. Readout errors correspond to the event of misidentifying the underlying state, meaning the state $\ket{0}$ yields output 1 or vice versa. While this misassignment is not perfectly symmetric in 0 and 1 in real hardware, approximating this error as a random bit flip usually serves as a good model. In order to measure the transformed Pauli strings $P_i(\gamma)$ when evaluating the energy cost $\loss_C(\gamma)$, we first append single-qubit gates to the noisy circuit $\tilde{A}(\vec{0})$ (including gate errors) to prepare the measurement basis. We then add random $X$ gates with probabilities $p_k$ on the qubits prior to the ideal readout, where the $p_k$ describe the misassignment errors on the qubits. Readout errors can be significantly larger than gate errors, making this addition essential to build a more accurate model.
\section{Methodology}
\label{sec:method}
\subsection{Selection of VQE problems as benchmarks}
In order to study the impact of Clapton we focus on a selection of physics and chemistry problem Hamiltonians.

\subsubsection{Physics models}
In statistical mechanics, two popular types of theoretical descriptions of magnetic systems are given by Ising models \cite{ising} and Heisenberg models \cite{heisenberg}. They are known to be able to show phase transitions in certain dimensions as well as complex non-classical ground states. Here we focus on the simple case of 1D models with constant couplings. The (transverse field) \emph{Ising model} is given by
\begin{equation}
    \ham = J \sum_{i=1}^{N-1} X_i X_{i+1} + \sum_{i=1}^N Z_i,
\end{equation}
which describes how $N$ spins (qubits) coupled in $x$-direction with strength $J$ interact with an external field of unit strength pointing in $z$-direction.

We further look at a special case of the (field-free) Heisenberg model, which couples the spins in all directions:
\begin{equation}
    \label{eq:xxz}
    \ham = \sum_{i=1}^{N-1} \qty(J X_i X_{i+1} + J Y_i Y_{i+1} + Z_i Z_{i+1}).
\end{equation}
Here the energy scale is set by choosing the $ZZ$-interaction to be of unit strength. Since the $XX$- and $YY$-interactions are constrained to have the same coupling $J$, this model is also referred to as the \emph{XXZ model}.

In this paper we study the cases $J=0.25, 0.50, 1.00$ for both the Ising model and the XXZ model.

\subsubsection{Chemistry models}
In order to evaluate Clapton for more complex Hamiltonians with a greater number of terms, we select three different molecules: H$_2$O, H$_6$ and LiH. We use Qiskit Nature~\cite{the_qiskit_nature_developers_and_contrib_2023_7828768} (which uses PySCF~\cite{pyscf} behind the scenes) to construct the corresponding Hamiltonians. They are computed in the STO-3G basis and mapped to qubits using the parity mapping~\cite{parity} with the two-qubit reduction applied. We limit the qubit count of all Hamiltonians to ten by restricting the active space of the molecules to six spatial orbitals. 
For each molecule we consider two different bond lengths $l$, one where classical methods yield high accuracy (small bond length) and one where the accuracy of classical methods is known to be less good (high bond length)~\cite{ravi2022cafqa}. The specific configurations are:
\begin{itemize}
    \item H$_2$O: $l = 1.0 \, \text{\AA}, 3.0 \, \text{\AA}$ -- 367 Hamiltonian terms
    \item H$_6$: $l = 1.0 \, \text{\AA}, 3.0 \, \text{\AA}$ -- 919 Hamiltonian terms
    \item LiH: $l = 1.5 \, \text{\AA}, 4.5 \, \text{\AA}$ -- 631 Hamiltonian terms
\end{itemize}



\subsection{Evaluation Method}
Since Clapton proposes a transformation scheme which is based on the initial point of a VQA, we compare its performance against the current state-of-the-art initialization method CAFQA. 
While the original CAFQA framework was implemented using Bayesian Optimization \cite{ravi2022cafqa}, we build our own adapted version that employs an optimization engine similar to the one shown Figure \ref{fig:clapton}. The main difference is that CAFQA optimizes the VQE ansatz $A$ over Clifford-compatible angles $\vec{\theta}$ and only considers a noiseless cost term similar to $\loss_0$.

Since CAFQA does not have a sense for noise, we furthermore implement an improved baseline to compare to: \emph{noise-aware CAFQA} (short: \emph{nCAFQA}). Just like CAFQA, nCAFQA finds an initialization for the original problem $\ham$ by searching over Clifford-compatible angles, but it uses the noise-equipped ansatz $\tilde{A}$. It computes cost terms similar to Clapton's $\loss_\noise$ and $\loss_0$ and also uses an adapted optimization engine. Comparing Clapton to nCAFQA will highlight the relevance of the Hamiltonian transformation step. While we refer to nCAFQA as a baseline, we would like to emphasize that it does not correspond to any prior art and gains its improvement over CAFQA through one of our major contributions, the classically efficient noise modeling.

We are comparing initilization schemes, thus we are mostly interested in the quality of the noisy evaluation of initial points. For Clapton this means $\theta = \vec{0}$ on the transformed problem and for CAFQA\slash nCAFQA $\theta = \theta_\text{CAFQA\slash nCAFQA}$ on the original problem. However, since for practical applications we care about the converged solutions after the application of VQE, we also run the full VQE for all problems to show the impact of the initialization schemes on convergence and the final result. We build the VQE simulation within the Qiskit framework and use the SPSA optimizer~\cite{spall1998overview}.

\subsubsection{Evaluation metric}
For all studies, our primary metric is given by the performance of Clapton relative to CAFQA\slash nCAFQA. We compute the \emph{relative improvement} $\eta$ which describes to what extent Clapton can reduce the gap to the problem's ideal ground state energy $E_0$ under noisy evaluation compared to a baseline:
\begin{equation}
    \eta = \frac{E_0 - E_\text{noisy}(\text{baseline})}{E_0 - E_\text{noisy}(\text{Clapton})}.
\end{equation}
Here ``baseline'' can either be CAFQA or nCAFQA. Since we consider benchmarks with up to ten qubits, it is possible to compute $E_0$ exactly by diagonalizing the Hamiltonian. As an example, $\eta = 2$ means that Clapton was able to reduce the gap by half.

\subsubsection{Evaluation on quantum backends}
\label{subsubsec:eval_devices}
For our main studies we use both realistic noise models (not Clifford-only simulable) and real quantum hardware to obtain energy values for our Hamiltonians of choice. To this end we use Qiskit~\cite{Qiskit} and its built-in simulators as well as the IBM Quantum Platform~\cite{IBMQS} to access real devices. In particular, we select noise model snapshots (known as fake backends) of the IBM machines \nairobi (seven qubits), \toronto (27 qubits) and \mumbai (27 qubits). We furthermore choose the cloud device \hanoi (27 qubits) for our quantum experiments. For \nairobi we constrain the set of benchmarks to only the physics models ($N=7$) while for the larger machines we investigate all models ($N=10$). Since quantum devices have a restricted topology, every VQE ansatz circuit $A$ needs to be mapped and routed on the machine before execution is possible. For Clapton this means that this so-called \emph{transpilation} step happens first to produce the transpiled ansatz $A'$ (now consistent with the hardware constraints), which is then fed to the Clapton scheme and used to evaluate the loss $\loss_\noise$ (Equation \eqref{eq:loss_n}). Clapton extracts the required parameters for gate and measurement errors from the noise models or machine calibration data accessible through Qiskit.

\subsubsection{Evaluation for specific sources of noise}
\label{subsubsec:specific_noise}
Additionally, for a subset of the $N=10$ benchmarks, we conduct studies for depolarizing noise and measurement noise individually, where we sweep a noise parameter $p$. In the case of depolarizing noise, $p$ represents the single-qubit gate error and ranges in $5 \cdot 10^{-4} \dots 5 \cdot 10^{-3}$. The two-qubit gate errors are set to $10p$. When studying measurement noise, $p$ describes the misassignment error (see Section \ref{subsubsec:meas_errors}) and we vary it within $5 \cdot 10^{-3} \dots 9.5 \cdot 10^{-2}$. For both studies we add thermal relaxation errors with characteristic decay times $T_1 = 50 \, \mu s, 100 \, \mu s, 150 \, \mu s$. The noise parameters are global across all qubits for this study. These chosen values are representative of what we find in the noise models of the quantum backends introduced above. In order to respect the realistic constraint of a limited hardware connectivity, we transpile the ten-qubit ansatz $A$ to match $\toronto$'s topology for optimization and evaluation. The latter is performed using Qiskit's AerSimulator.

\section{Evaluation}
\label{sec:eval}
\subsection{VQE initial and final points on backends}
\begin{figure*}[htbp]
    \centering
    \includegraphics[scale=1.0]{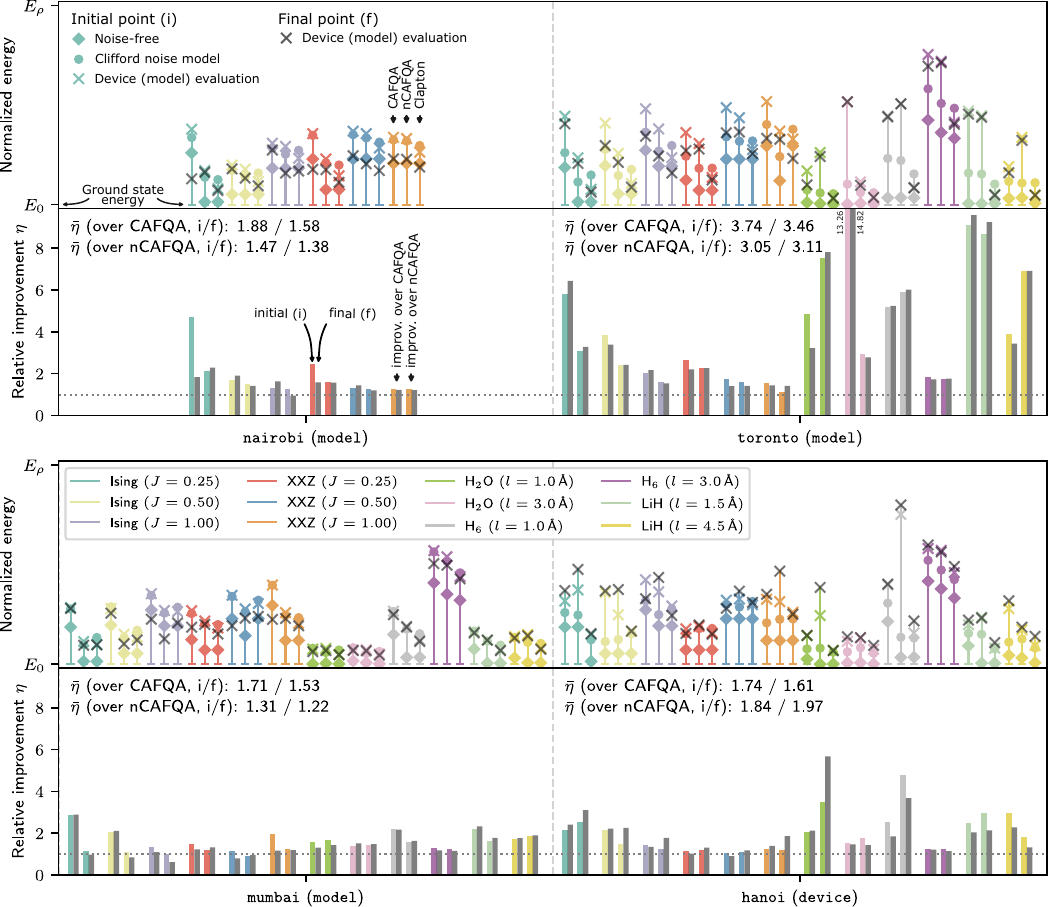}
    \caption{Comparing the VQE initialization quality of Clapton against the baselines CAFQA and nCAFQA, as measured by achieved energy and relative improvement $\eta$, across a variety of benchmarks (color-coded). Results are shown for three noise models (\nairobi\!, \toronto\!, \mumbai\!) and one real quantum device (\hanoi\!). Top rows: Energy values of found solutions for each method on an absolute, normalized scale (see main text). Different symbols refer to the evaluation of each solution in different noise environments. Energies under full model/device evaluation ($\times$) are the most relevant. The dark $\times$ refers to the energy after a couple hundred VQE iterations (final point (f)) when starting from the colored $\times$ (initial point (i)). Bottom rows: Relative improvements of Clapton over CAFQA and nCAFQA, evaluated for the initial $\times$ and the final $\times$. The inset text shows the geometric mean $\bar{\eta}$ across all benchmarks.}
    \label{fig:all_points}
\end{figure*}
For the general evaluation of our scheme under realistic noise effects, we start by running Clapton, nCAFQA and CAFQA for all VQE problems on all the backends as described in Section \ref{subsubsec:eval_devices}. 
In order to find the quality of the initial VQE points, we need to evaluate the realistic energy estimates for the specific problems and inputs. We achieve this by using Qiskit's built-in simulators for the fake backends (``models'') \nairobi\!, \toronto and \mumbai\!, while for the real machine \hanoi we send jobs to the IBM Quantum Platform. Afterwards we run up to a thousand VQE iterations from each initial point on those backend noise models and find the quality of the converged solution.  Therefore we only evaluate the final points on the actual hardware. Finally we determine relative improvements $\eta$ for initial and final points and combine all findings in Figure \ref{fig:all_points}.

Figure \ref{fig:all_points} consists of two similar parts showing results for two backends respectfully. The top rows present the quality analysis of found solutions from each method for each benchmark. Different colors correspond to different benchmarks. For easier visualization they are normalized such that two fixpoints align: the ground state energy $E_0$ and the energy of the fully mixed state $E_\rho = \tr[\rho \ham] = \tr[\ham]/2^N$ ($\rho = I_N/2^N$). We show three different energy values for each initial point: (1) The noiseless case ($\diamond$), which is a lower bound for any noisy value and used by all three methods in their optimization (Clapton: $\loss_0$). (2) The energy under the approximate Clifford noise model ($\circ$), which Clapton and nCAFQA use to evaluate the other part of their cost function (Clapton: $\loss_\noise$). (3) The energy under the full complex noise model derived from a fake backend or, in the case of \hanoi\!, the result from the hardware experiment ($\times$). Clearly the most relevant property is (3). We observe that for almost all benchmarks Clapton finds significantly better starting points than the other two methods, while nCAFQA often performs better than CAFQA as one would expect. Another interesting observation is that the discrepancy between (2) and (3) is typically the lowest for Clapton (see also Figure \ref{fig:clapton_mini} for a magnified version of one benchmark), indicating that the Clifford noise modeling becomes more accurate when we transform to the $\ket{0}$ state. This aligns with the Clapton hypothesis that exploring around the $\ket{0}$ is favorable as the influence of relaxation (which is non-Clifford) is mitigated.

We mark the final point after the application of VQE with a dark $\times$ and find that the advantage of Clapton persists through the convergence in nearly all cases. In few cases a CAFQA or nCAFQA instance is able to catch up and overtake a Clapton instance. We will address this further in the next section. In the case of \hanoi these results were obtained by extracting the noise model from calibration data and perform the VQE iterations on the classical simulator. The final points are then evaluated using hardware experiments. The overhead of submitting jobs to the cloud for every iteration in combination with long queue times would render full VQE on the real devices infeasible.

The bottom rows visualize the relative improvements that Clapton enables, both for the initial as well as the final points. When comparing to CAFQA, we find (geometric) mean improvements ranging from 1.7x to 3.7x for initial points, with improvements for individual VQE problems up to 13.3x. For the final points the mean improvements range 1.5x to 3.5x (14.8x max). As expected, the improvements over nCAFQA are smaller but still significant. This is because nCAFQA already benefits strongly from modeling gate and readout errors during its optimization which is one of our key contributions. The Clapton transformation then further adds resilience to state decay which yields its advantage over nCAFQA. However, one should keep in mind that even a relative improvement of (for example) 1.3 is remarkable as it corresponds to a $\sim \! 23\%$ gap reduction.

Looking across devices we observe that experiments executed on the \toronto model particularly benefit from Clapton, while looking across problem types shows that chemistry problems profit greatly. We suspect the latter to be caused by the substantially larger number of Pauli terms in the respective Hamiltonians, allowing for a greater impact of the Clapton transformation -- this is promising for real world applications. We emphasize the observed improvements for the real quantum device \hanoi\!: Device execution of noise-model-optimized transformations show improvements across all benchmarks except for one case where the final point was slightly worse. This should not be taken for granted as there can be no guarantee that simplified noise models represent effects on real hardware well.

\begin{figure*}[htbp]
    \centering
    \includegraphics[scale=1.0]{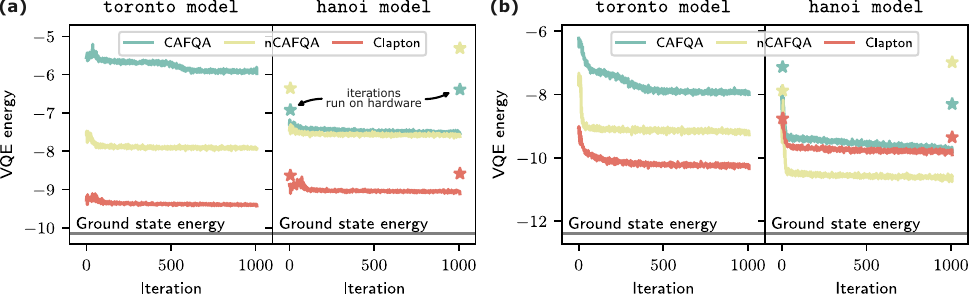}
    \caption{Visualizations of the VQE convergence of the ten-qubit XXZ model with (a) $J=0.25$ and (b) $J=1.00$, run on two different noise models. The Clapton initialization starts better than the baselines and typically converges better. Initial and final evaluations on real machine \hanoi are shown with stars. Hardware experiments yield worse final points in some cases, suggesting a device-model discrepancy that Clapton does not seem to be strongly affected by.}
    \label{fig:convergence}
\end{figure*}

\subsubsection{VQE convergence}
In Figure \ref{fig:convergence}, we visualize the VQE convergence of the ten-qubit XXZ model with (a) $J=0.25$ and (b) $J=1.00$ when run on the models of the four 27-qubit machines \toronto and \hanoi\!. $J=0.25$ represents a case where stabilizer states can provide a good approximation to the ground state while for $J=1.00$ that is not the case. We see here that Clapton finds good starting points and converges well in both cases. We note that even though the Clapton solution starts better than the nCAFQA solution for $J=1.00$ on \hanoi\!, the latter overtakes quickly and finds a better final point. This suggests that the cause for these two observation lies with SPSA and its configuration. Nondeterminism in noise, barren plateaus and uncertainty in sampled energy estimates can be detrimental to its performance. This issue could be addressed by configuring the optimizer hyperparameters more carefully or studying other optimization frameworks, which is outside the scope of this work. We emphasize that while Clapton is able to produce high-quality starting points, it does not come with the guarantee of best convergence every time, as this is influenced by factors  Clapton (and any other VQA initialization method) does not have control over. 

The experimental results of the initial and final points on \hanoi are visualized by stars. It is interesting to observe that in some cases experiments on real hardware yield worse final points that are worse than the initial points. This suggests a discrepancy between the extracted noise model which is used for the simulated VQE and the actual device. This is unsurprising as accurately capturing the quantum effects on many-qubit devices is very challenging. We note that even though Clapton did not find the best final point for $J=1.00$ on \hanoi according to the simulation, it achieved the best experimental result.

\subsection{Varying the strength of individual noise channels}
Having presented the positive impact on Clapton when applied to full machine noise models, we now turn to studying individual noise channels in an isolated environment in particular. As described in Section \ref{subsubsec:specific_noise}, we sweep gate errors and measurement errors, which are both essential Clapton ingredients, for different decay times $T_1$. Results are presented for the initial point of selected VQE problems and we focus on comparing to the modified baseline nCAFQA.
\begin{figure}[htbp]
    \centering
    \includegraphics[scale=1.0]{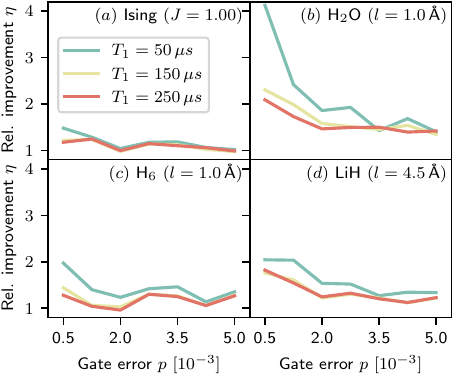}
    \caption{The impact of Clapton when solely sweeping the single-qubit gate error $p$ (two-qubit gate error $10p$), for three different decay times and a selection of benchmarks. Relative improvement is evaluated at the initial VQE point.}
    \label{fig:gate_errors}
\end{figure}

In Figure \ref{fig:gate_errors} we show the relative improvements for four different benchmarks when sweeping the gate errors. We note that the Clapton's transformation is especially useful for lower error rates when solving the complex chemistry models, while we observe an upwards trend for the Ising model as the gate error increases. As expected, shorter $T_1$ times, meaning stronger thermal relaxation, typically lead to better improvements as the nCAFQA initial point suffers from more state decays. This influence is actively counteracted by Clapton by transforming to the qubit ground state $\ket{0}$.

\begin{figure}[htbp]
    \centering
    \includegraphics[scale=1.0]{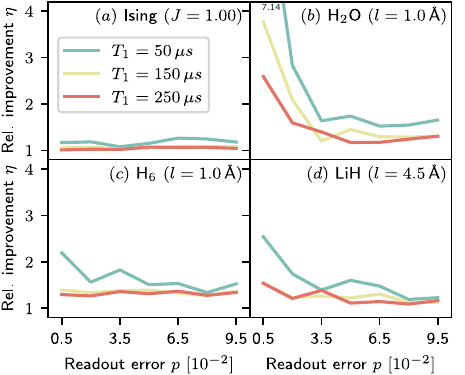}
    \caption{The impact of Clapton when solely sweeping the measurement error $p$, for three different decay times and a selection of benchmarks. Relative improvement is evaluated at the initial VQE point.}
    \label{fig:meas_errors}
\end{figure}

We find similar results when varying measurement errors on the readout qubits as shown in Figure \ref{fig:meas_errors}. The noticable difference is that the Ising model seems to be more robust to this kind of error as only small improvements over the baseline can be observed. The chemistry problems still profit significantly from the transformation, underlining that Clapton is well suited for this type of Hamiltonian structure.

\begin{figure*}[htbp]
    \centering
    \includegraphics[scale=1.0]{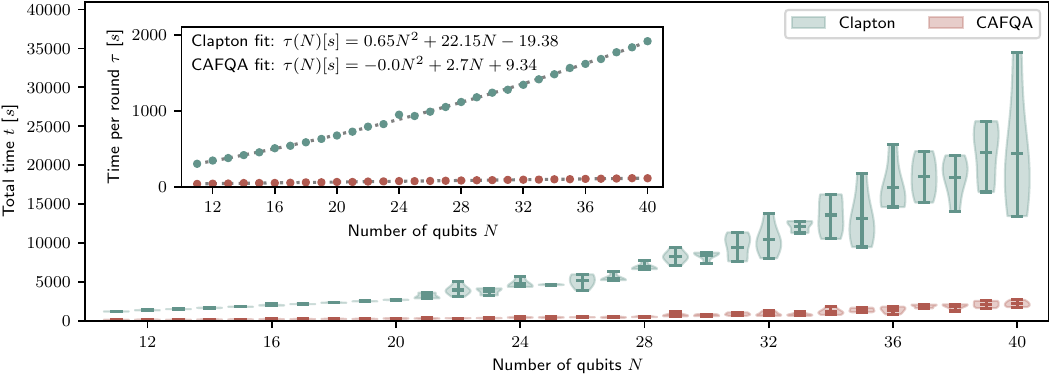}
    \caption{Scaling of Clapton optimization time with the number of qubits $N$ for the Ising model ($J=0.25$). Each instance is run from five initial guesses using $\nstarts = 10$. The inset presents the scaling of the average optimization time per round, which follows an $\mathcal{O}(N^2)$ trend. The baseline CAFQA method shows a considerably faster performance as it only computes noiseless evaluations.}
    \label{fig:scaling}
\end{figure*}
\subsection{Clapton's performance scaling}
In this last section we study the classical compute Clapton requires for its optimization. We investigate the performance scaling with respect to the qubit count $N$ and to this end consider the Ising model with $\mathcal{O}(N)$ terms. For the purpose of this study transpilation is not required so we keep the ansatz structure $A$, which grows in depth and width with $\mathcal{O}(N)$. Thus the complexity of the circuit that needs to be simulated scales with $\mathcal{O}(N^2)$.

For $N=11, \dots, 40$ we run Clapton five times from random initial configurations and measure the time until convergence (that is, no loss function improvement even after two retry rounds). We specify $\nstarts = 10$ and use ten processes, thus here we only parallelize over the genetic algorithm instances $\GA{1}, \dots, \GA{10}$ and not over the Pauli terms or population size.

Figure \ref{fig:scaling} shows the scaling of the total compute time $t$ with qubit count $N$. Up to 20 qubits we barely observe any variation in optimization time across the five guesses, noting an essentially linear increase with $N$. For higher qubit counts it shows that an increasing number of rounds is required to achieve convergence, which is expected due to the growing number of parameters ($\mathcal{O}(N)$) and the vast search space $\Gamma$ (dimension exponential in $N$). While the right half of Figure \ref{fig:scaling} hints at this exponential scaling, it is important to note that the compute time never exceeded ten hours even for the $N=40$ case, which is already well beyond what would be classically simulable if we considered more accurate quantum models.

The inset visualizes the average time per round $\tau$, which scales like the circuit spacetime volume with $\mathcal{O}(N)$. This makes sense as rounds have a fixed number of iterations and their speed is therefore determined by stim's simulation efficiency. The round duration could be reduced by a constant factor by furthermore parallelizing over the Pauli term count and \GA{} population size in each iteration, which would speed up Clapton's performance by a constant factor. This can be realized in a straightforward way due to the parallel nature of the Clapton design with the use of genetic algorithms (see Figure \ref{fig:clapton}), underlining Clapton's efficient and scalable implementation.

We perform the same study for the baseline methods CAFQA and nCAFQA. While nCAFQA yields a performance scaling that is essentially equivalent to Clapton due to its very similar design (we thus do not show it in Figure \ref{fig:scaling}), our CAFQA implementation runs in much less time to achieve convergence. This is because only noiseless Clifford simulations are required where no sampling is necessary, every Pauli expectation value can be determined in one circuit evaluation. Figure \ref{fig:scaling} includes these results and the inset shows how stim's fast Clifford simulation leads to linear scaling of round time $\tau$ over the considered qubit count range.

This study as well as all other Clapton or CAFQA optimizations demonstrated in this paper were performed on Intel Gold 6248R CPUs.

\section{Related work}
\label{sec:related_work}
Although Clifford group operations and Pauli measurements do not suffice for universal quantum computing, many quantum domains such as quantum networks \cite{Veitch2014}, error correction codes \cite{QECIntro}, teleportation \cite{gottesman1999demonstrating} and error mitigation \cite{czarnik2020error, strikis2021learningbased} focus on applications within the Clifford space.

In terms of related work in the space of VQA optimization,
Mitarai et al. \cite{mitarai2020quadratic} advocate a perturbative expansion of the cost function from Hartree-Fock (HF) initialization, potentially yielding improved performance. Sauvage et al. \cite{sauvage2021flip} introduce FLIP using machine learning for initialization. MetaVQE by Cervera et al. \cite{Cervera2021} encodes Hamiltonian parameters into the ansatz for a more reliable estimate of the ground state. Wurtz et al. \cite{wurtz2021classically} use classical solutions to aid some special sets of problem-specific circuits. 
Wang et al. propose QuantumNAS~\cite{wang2022quantumnas}, a framework for noise-adaptive co-search of the variational circuit structure and qubit mapping. Their approach is substantially different to the Clapton proposal as they leave the degrees of freedom that come from transforming the Hamiltonian untouched and focus on the manipulation of the original problem implementation.

Various orthogonal error mitigation strategies are also available to address quantum errors at different stages of a quantum program: Pre-processing techniques (e.g. noise-aware mapping \cite{murali2019noise}), post-processing techniques (e.g. measurement error mitigation \cite{tannu2019mitigating}) or in-situ techniques (e.g. dynamical decoupling \cite{DDBiercuk_2011,jurcevic2021demonstration,DDKhodjasteh_2007,DDPokharel_2018}).
Previous studies \cite{kandala2019error, wang2021can} emphasize error mitigation's role in enhancing VQA accuracy and trainability, with techniques like zero-noise extrapolation (ZNE) and addressing noise-induced barren plateaus. The VAQEM approach \cite{ravi2021vaqem} tailors existing error mitigation techniques dynamically to real-time execution characteristics of VQAs on a quantum machine.

\section{Conclusion}
In this current NISQ era VQAs represent a promising avenue to obtain useful insights from quantum devices. However, noise is the defining characteristic of this period and will always have a detrimental effect on any quantum computing results, rendering adjustment to this noise through sophisticated error mitigation techniques indispensable. The Clapton proposal directly addresses this issue for VQAs by exploiting the freedom that can be found in the problem formulation (the Hamiltonian $\ham$). It makes use of the observation that, given a specific quantum device, some Hamiltonians are less susceptible to its noise pattern than others when running the VQA. Clapton performs a classically simulable search over transformations that make the problem formulation more compatible with the hardware noise characteristics, yielding a new Hamiltonian $\hat\ham$ that is then used during the full VQA execution. This work shows the beneficial impact of this approach across a variety of benchmarks. 

Clapton can be viewed as a pre-processing error mitigation technique, which may be combined with other popular error mitigation methods outlined in Section \ref{sec:related_work}; an interesting idea for future work. It is a pure software tool which models and simulates the target quantum machine efficiently by expressing all operations in terms of Clifford gates.

As quantum computing research moves closer to fault-tolerance and error correction, it is important to point out that the Clapton is not limited to the NISQ regime. Error-corrected quantum devices will also run VQEs and errors on (logical) qubits will still occur, albeit with a significantly reduced rate. In fact, errors on these machines are discretized and typically expressed in terms of bit flips ($X$ operations) and phase flips ($Z$ operations) \cite{shor_ftqc}, which directly suggests the depolarizing error model to describe these effects. Therefore, given the implementation of Clapton, it might prove itself to be even more relevant and accurate in the future. 

\begin{acks}
This work is funded in part by EPiQC, an NSF Expedition in Computing, under award CCF-1730449; in part by STAQ under award NSF Phy-1818914/232580; in part by the US Department of Energy Office of Advanced Scientific Computing Research, Accelerated 
Research for Quantum Computing Program; and in part by the NSF Quantum Leap Challenge Institute for Hybrid Quantum Architectures and Networks (NSF Award 2016136), in part based upon work supported by the U.S. Department of Energy, Office of Science, National Quantum 
Information Science Research Centers, and in part by the Army Research Office under Grant Number W911NF-23-1-0077. The views and conclusions contained in this document are those of the authors and should not be interpreted as representing the official policies, either expressed or implied, of the U.S. Government. The U.S. Government is authorized to reproduce and distribute reprints for Government purposes notwithstanding any copyright notation herein.

FTC is the Chief Scientist for Quantum Software at Infleqtion and an advisor to Quantum Circuits, Inc.

This research used resources of the Oak Ridge Leadership Computing Facility, which is a DOE Office of Science User Facility supported under Contract DE-AC05-00OR22725.

This work was completed in part with resources provided by the University of Chicago’s Research Computing Center.
\end{acks}

\bibliographystyle{plain}
\bibliography{ref}

\end{document}